\documentclass[12pt]{article}
\usepackage{axodraw}
\usepackage{amssymb}
\usepackage{graphicx}

\begin{document}

\parskip=0.3cm
\begin{titlepage}

\hfill \vbox{\hbox{DFPD 01/TH/44}\hbox{DFCAL-TH 01/6}\hbox{BITP-01-20}
\hbox{November 2001}}

\vskip 0.5cm

\centerline{\bf $J/\psi$ PHOTOPRODUCTION AT HERA}

\vskip 0.3cm

\centerline{R.~Fiore$^{a\dagger}$, L.L.~Jenkovszky$^{b\ddagger}$,
F.~Paccanoni$^{c\ast}$, A.~Papa$^{a\dagger}$}

\vskip 0.1cm

\centerline{$^{a}$ \sl  Dipartimento di Fisica, Universit\`a della Calabria}
\centerline{\sl Istituto Nazionale di Fisica Nucleare, Gruppo collegato di Cosenza}
\centerline{\sl I-87036 Arcavacata di Rende, Cosenza, Italy} 

\centerline{$^{b}$ \sl  Bogolyubov Institute for Theoretical Physics}
\centerline{\sl Academy of Sciences of Ukraine}
\centerline{\sl UA-03143 Kiev, Ukraine}

\centerline{$^{c}$ \sl Dipartimento di Fisica, Universit\`a di Padova}
\centerline{\sl Istituto Nazionale di Fisica Nucleare, Sezione di Padova}
\centerline{\sl via F. Marzolo 8, I-35131 Padova, Italy}

\vskip 0.1cm

\begin{abstract}
We revise and update an earlier model for $J/\psi$ photoproduction 
based on a dipole Pomeron exchange. We show that the H1 and ZEUS
experimental data reported recently can be well fitted 
by a soft Pomeron alone.   

\end{abstract}

\vskip 0.1cm

\vfill

\hrule

$
\begin{array}{ll}
^{\dagger}\mbox{{\it e-mail address:}} &
   \mbox{FIORE,~PAPA@CS.INFN.IT} \\
^{\ddagger}\mbox{{\it e-mail address:}} &
\mbox{JENK@GLUK.ORG} \\
^{\ast}\mbox{{\it e-mail address:}} &
   \mbox{PACCANONI@PD.INFN.IT}
\end{array}
$

\end{titlepage}
\eject
\newpage

In this paper we explore further the idea about the ``softness'' of heavy
vector meson photoproduction put forward in a previous publication~\cite{FJP}. 
The basic assumptions are:

1. Photoproduction of heavy vector mesons is a soft process.
It can be described by the exchange of either a supercritical Pomeron with a 
low intercept, typically $\alpha_{I\!\!P}(0)-1\approx 0.08,$ or a dipole one with a 
unit intercept. We stick to the second possibility since it gives better fits 
to the data. We remind the reader that the dipole Pomeron produces 
(logarithmically) rising cross sections even at a unity intercept of the 
trajectory, $\alpha_{I\!\!P}(t=0)=1$ (see Refs.~\cite{FJP,JMP} and references 
therein).

2. The Pomeron -- whatever it be -- contains more then just a single (and thus
factorisable) term. The simplest and natural choice is a sum of a constant and
a (moderately) rising term. Their interference leads~\cite{FJP,JMP} to the 
delay in reaching the asymptotics of $\sigma_{el}(s)$, observed at HERA. In other 
words, the relatively rapid increase of $\sigma_{el}(s)$ in the intermediate 
energy region of HERA is a transitory effect, followed by a subsequent 
slow-down. Of course, the presence of more than one term in the Pomeron 
will break factorisation (to be restored asymptotically).

3. The $\gamma {I\!\!P}V$ vertex, shown as $\beta_2(t)$ in Fig.~1, may have a 
$t$-dependence different from that in a typical hadronic vertex. This deviation 
reflects of the departure from vector meson dominance in the case of heavy 
mesons. Following the arguments presented in Ref.~\cite{FJP}, we add in 
this vertex a term proportional to $t$ and thus vanishing towards $t=0$. Its 
presence can make  the determination of the slope parameter somewhat 
independent of the actual value of the total and differential cross 
section in the forward direction.
The role of this term will be determined from the fits to the data.  

\begin{figure}
\begin{center}
\begin{picture}(200,200)(35,-40)
\Text(50,142)[]{$q$}
\Text(190,142)[]{$p_V$}
\Photon(40,140)(101,105){2}{6}
\ArrowLine(75,122)(77,120.5)
\GCirc(120,100){20}{1}
\Text(120,100)[]{$\beta_2(t)$}
\Line(139,102.5)(200.3,139.6)
\Line(139,103.5)(200,140.5)
\ArrowLine(170,121.5)(172,122.8)
\ZigZag(120,80)(120,5){3}{8}
\Text(145,43)[]{$\alpha_{I\!\!P}(t)$}
\ArrowLine(40,-35)(120,5)
\ArrowLine(120,5)(200,-35)
\Text(120,-11)[]{$\beta_1(t)$}
\Text(50,-39)[]{$P$}
\Text(190,-39)[]{$P'$}
\end{picture}
\end{center}
\caption[]{\small Elastic photoproduction with an inelastic
$\gamma {I\!\!P}V$ vertex. The wiggle line represents the dipole Pomeron.
The diagram corresponds to the sum of two diagrams, i.e. one with a simple and the
other with a double pole exchange.}
\label{fig1}
\end{figure}
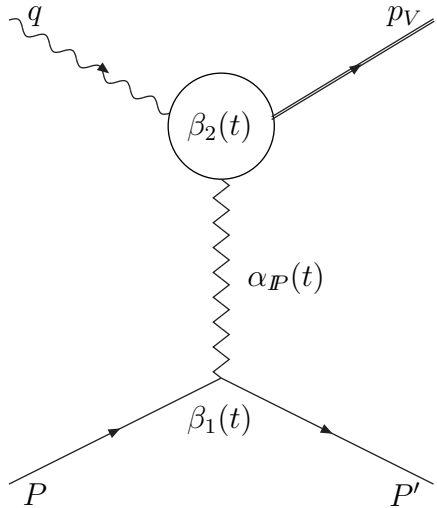

4. The Pomeron trajectory is not a linear function. This is a direct  
consequence of the theory (analyticity and unitarity) and it has been verified 
in many reactions -- elastic and inelastic~\cite{Brandt}. A reasonable model for
the trajectory, compatible with the asymptotic bounds, yet feasible 
phenomenologically, is~\cite{FJMPP} 
\begin{equation}
\alpha_{I\!\!P}(t)=1+\gamma(\sqrt{t_0}-\sqrt{t_0-t})~.
\end{equation}

In the present paper we concentrate on $J/\psi$ photoproduction in the HERA
kinematical region. This process is particularly advantageous in checking 
Pomeron models since, on one hand, due to the OZI rule~\cite{OZI},
contributions from secondary trajectories here can be completely neglected, and,
on the other hand, a lot of experimental data has been accumulated -- much more 
than e.g. in the case of $\Upsilon$ photoproduction. This is the main reason 
why this reaction receives recently so much attention~\cite{Levin}. 

We use the following notation: the square of the center of mass system (c.m.s.) 
energy and the momentum transfer to the proton are, respectively,
\begin{equation}
W^2=s=(q+P)^2, \quad  \quad \quad  t=(P-P')^2
\end{equation}
with 
\begin{equation}
|t|_{min}\approx m_p^2{(M_V^2+Q^2)^2\over{W^4}}.
\end{equation}
Here $M_V$ is the vector-meson mass, $m_p$ is the proton mass and $Q^2=-q^2$ is the 
photon virtuality. Let us note that at HERA one has 
20 GeV $<W<240$ GeV, $-13$ GeV$^2<t<- 10^{-4}$ GeV$^2$.

We shall be particularly interested in the interpretation of the latest, 
highest-energy data, those published by the H1 Collaboration~\cite{Adlo,Aid}
and those reported by the ZEUS Collaboration at the 2000 Osaka 
Conference~\cite{Osaka}. In our opinion these new data may become critical 
in discriminating between models for the Pomeron; in particular, they may be 
indicative of the presence of any hard component in $J/\psi$ 
photoproduction~\cite{DL}.

Following these preliminary remarks, we write the scattering amplitude for 
the reaction $\gamma \: p\rightarrow J/\psi \: p$ as    
\begin{equation}
A(s,t)\propto \biggl[a\exp {(bt)}+ct\exp{(dt)}\biggr]
\left(\frac{s}{s_0}\right)^{\alpha_{I\!\!P}(t)}
\left[\ln\left(\frac{s}{s_0}\right)+g-{i\pi\over 2}\right]\xi(\alpha_{I\!\!P}(t))~, 
\end{equation}
where $\xi(\alpha_{I\!\!P}(t))=$exp$(-i\pi\alpha_{I\!\!P}(t)$/2).
In Eq.~(4), the residue (the first factor in the r.h.s.) is a generalization of 
that used
in Ref.~\cite{FJP}, $\exp{(bt)}(1+ct)$. Actually, this residue is not
exactly the product of the two vertices $\beta_1(t)$ and $\beta_2(t)$ 
(see Fig.~1), since the amplitude is a sum of two terms, each with its 
vertices. The second parentheses contains terms 
typical of a dipole Pomeron (see Refs.~\cite{FJP,JMP} and references therein): a 
constant, a logarithmically rising one and an imaginary part, coming from the 
signature factor. We are using a Pomeron trajectory of the form (1) with 
$t_0=4m_{\pi}^2$ and set $\gamma=m_{\pi}$ to get the ``standard'' value for 
the Pomeron slope, $\alpha_{I\!\!P}^\prime(0)\approx 0.25\;$ GeV$^{-2}$.

We have calculated the elastic differential cross 
section of $J/\psi$ photoproduction according to the formula
\begin{equation}
{d\sigma\over{dt}}=\biggl[a\exp{(bt)}+ct\exp{(dt)}\biggr]^2
\left(\frac{s}{s_0}\right)^{2\alpha_{I\!\!P}(t)-2}
\left[\left(\ln\left(\frac{s}{s_0}\right)+g\right)^2+{\pi ^2\over 4}\right] 
\end{equation}
and have fitted it to the recent H1~\cite{Adlo,Aid} and ZEUS~\cite{Priv} data.

With the parameters obtained from the fits to the differential cross section, the 
integrated elastic cross section 
\begin{equation}
\sigma_{el}(s)=\int_{t_+}^{t_-} dt {d\sigma\over {dt}}~, 
\end{equation}
where $t_-=O[1/s^2]\sim 0$ and $t_+\sim -s$, as well as the local
slope of the differential cross section for various fixed values of $t$,
\begin{equation}
B(s)=\frac{d}{dt} \left (\ln {d\sigma\over{dt}}\right)~,
\end{equation}
were calculated. 

Setting $s_0=1$ GeV$^2$, we fitted Eq.~(5) separately to the 
H1~\cite{Adlo,Aid} data and the preliminary 
ZEUS data~\cite{Priv}, presented at the $2000$ Osaka Conference~\cite{Osaka},
since the two sets of the data differ substantially. 

When fitting Eq.~(5) to the H1 data, we found a wide
region in the space of parameters where the $\chi^2$/d.o.f. is 
lower than 1. In order to remedy this ambiguity, we have chosen
a set of parameters which leads to a $\sigma_{el}(s)$ close to the 
data points quoted by the H1 Collaboration~\cite{Adlo,Aid}. For
this purpose, we were constrained to fix a subset of parameters, namely 
$a$, $b$ and $c$, and to leave free the remaining two. In this way 
we found a minimum with $\chi^2$/d.o.f=0.96. We remind, however, that 
$\sigma_{el}(s)$ does not result from direct measurements, but is 
always model-dependent.

The values of the parameters are: $a=5.203$ nb$^{1/2}$ GeV$^{-1}$, 
$b=2.086$ GeV$^{-2}$, $c=-2.838$ nb$^{1/2}$ GeV$^{-2}$, 
$d=(2.343\pm 0.351)$ GeV$^{-2}$ and $g=-5.736\pm 0.100$.
Fig.~2 shows the curve for ${d\sigma/dt}$ resulting from the fit, 
together with the H1 data points~\cite{Adlo}.

\begin{figure}[tb]
\begin{center}
\includegraphics[scale=0.77]{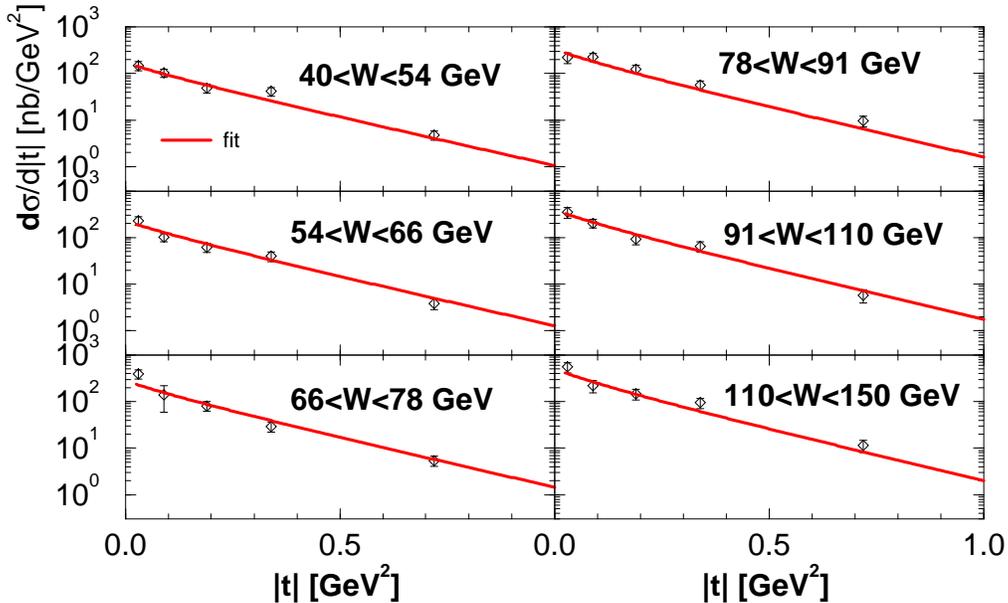}
\end{center}
\caption[]{\small Differential cross sections $d\sigma/dt$ [nb/GeV$^2$]
for elastic $J/\psi$ photoproduction for different bins of $W$. Data are from
H1~\cite{Adlo}. The solid lines represent the result of the fit.}
\label{fig2}
\end{figure}

\begin{figure}[tb]
\begin{center}
\includegraphics[scale=0.68]{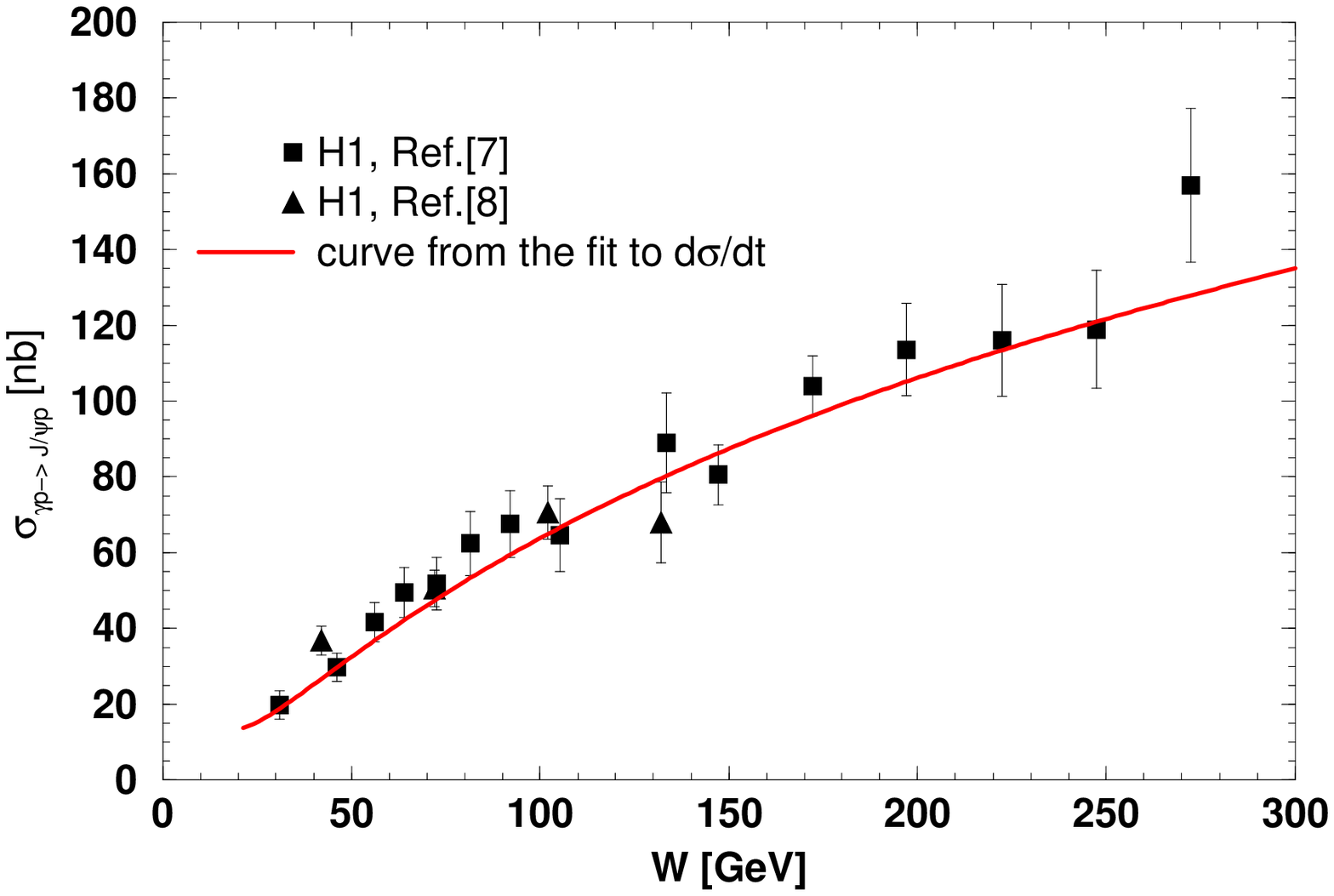}
\end{center}
\caption[]{\small Elastic $J/\psi$ photoproduction cross section [nb].
Data are from H1~\cite{Adlo,Aid}. The solid line represents the result of the fit
to the differential cross sections $d\sigma/dt$.}
\label{fig3}
\end{figure}

The elastic cross section $\sigma_{el}(s),$ obtained from our model, is compared 
in Fig.~3 with the H1 data points~\cite{Adlo,Aid}. Note that the high-energy part 
of the theoretical curve does not tend to ``harden'' (the rate of its rise is even 
slowing down).

Our fit to the ZEUS data on ${d\sigma/dt}$~\cite{Priv}, on the other hand, 
is unambiguous.  After noticing that $d$ varies little in the fit, we have 
fixed its value, thus leaving only four parameters free.    
We set $d=0.851$ GeV$^{-2}$, the values of the fitted parameters being
$a=(3.856\pm 0.174)$ nb$^{1/2}$ GeV$^{-1}$, $b=(1.625\pm 0.091)$ 
GeV$^{-2}$, $c=(-0.936\pm 0.144)$ nb$^{1/2}$ GeV$^{-2}$, $g=-4.126\pm 0.126$.
The fit leads to a single, pronounced minimum with $\chi^2$/d.o.f. = 1.04 
and the resulting curves for ${d\sigma/dt}$, for different bins of $W$, are shown 
in Fig.~4. This figure should be compared with Fig.~5 of Ref.~\cite{Osaka}.

\begin{figure}[tb]
\begin{center}
\includegraphics[scale=0.75]{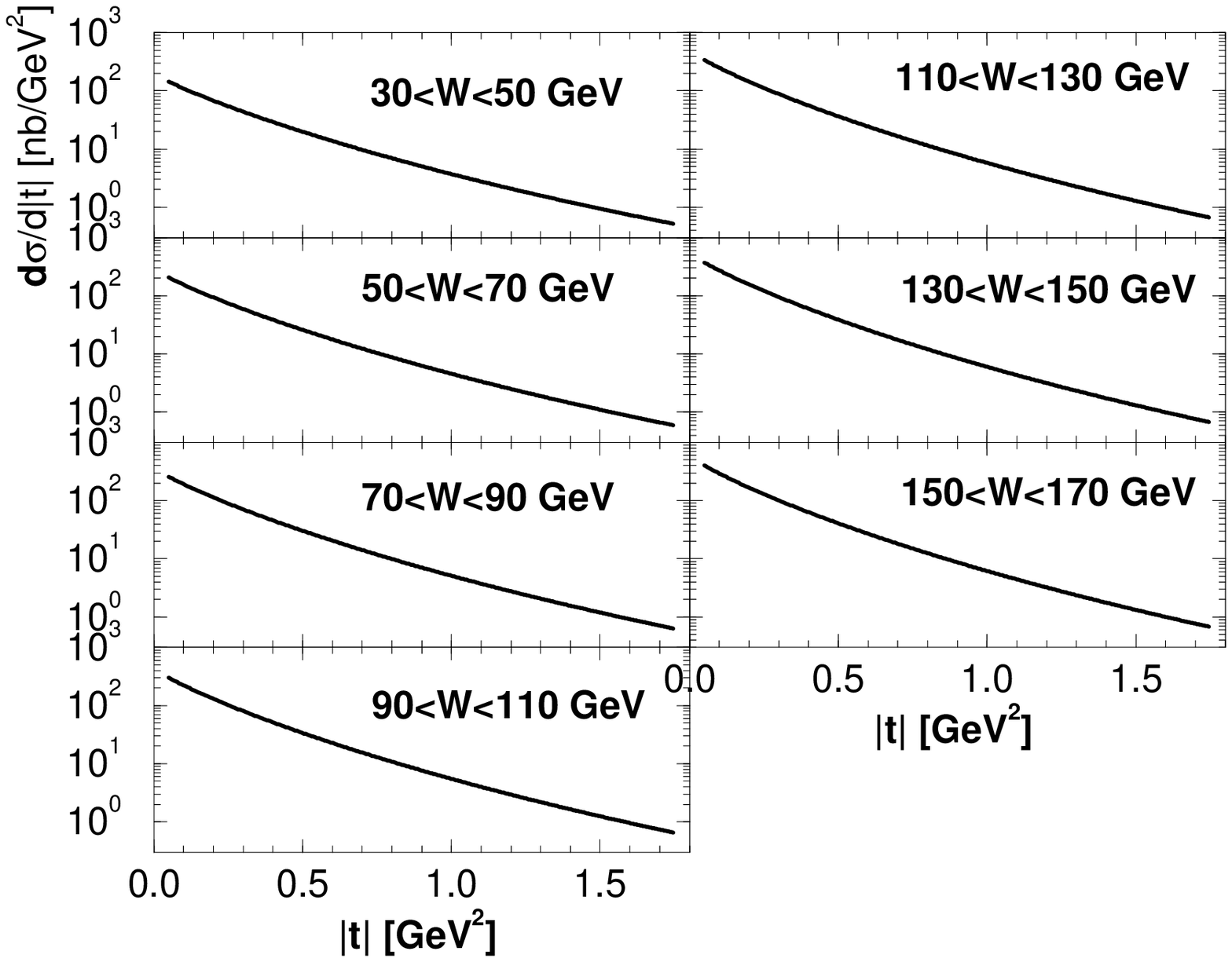}
\end{center}
\caption[]{\small The solid lines represent the result of the fit
to the ZEUS data~\cite{Priv} for the differential cross sections $d\sigma/dt$ 
[nb/GeV$^2$] for elastic $J/\psi$ photoproduction, for different bins of $W$.}
\label{fig4}
\end{figure}

\begin{figure}[tb]
\begin{center}
\includegraphics[scale=0.68]{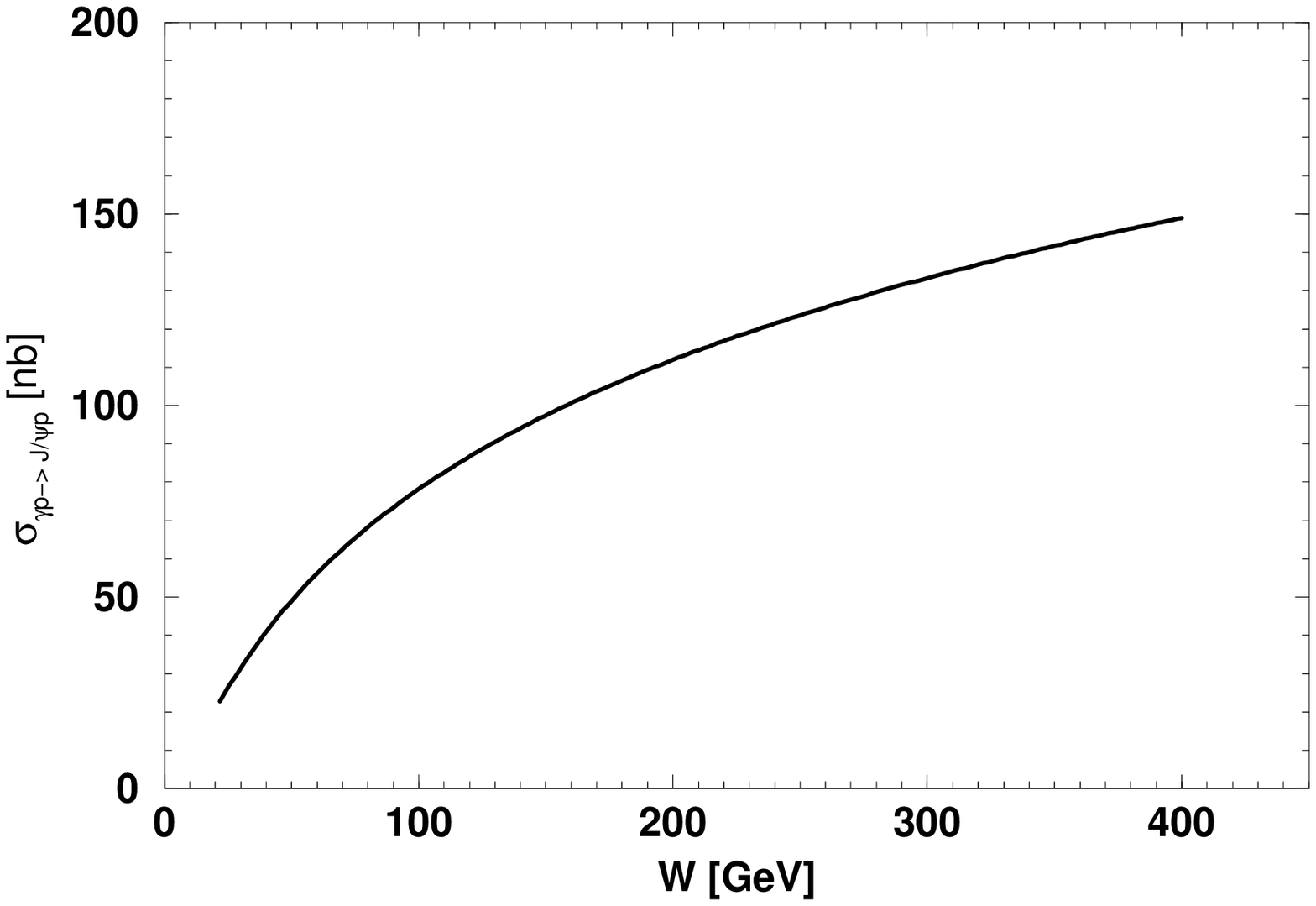}
\end{center}
\caption[]{\small The solid line represents the elastic $J/\psi$ photoproduction 
cross section [nb], resulting from the fit to the ZEUS data~\cite{Priv} 
for ${d\sigma/dt}$.}
\label{fig5}
\end{figure}

With the parameters obtained from the fit to the ZEUS data~\cite{Priv} on the 
differential cross section, the elastic cross section, $\sigma_{el}(s)$ has 
been calculated according to Eq.~(6). The result is presented in Fig.~5.
A comparison with Fig.~4 of Ref.~\cite{Osaka} shows a very good agreement
with the ZEUS data for $\sigma_{el}(s)$.

Finally, the behavior of the local slope $B(s)$, calculated from Eq.~(7) 
as a function of $s$ for various fixed values of $t$, is shown in Fig.~6.
As expected from the differential cross section and the curvature of the Pomeron 
trajectory~(1), the local slope decreases with $|t|$. Its value at 
$t\approx -0.3$ GeV$^2$ meets the experimental measurements
(see Fig.~6 of Ref.~{\cite{Osaka}). This is
quite understandable, since the experimental value is the average over a wide 
interval in $t$ covering the measurements. The rise of the slope towards $t=0$ is a 
well-known phenomenon in hadronic physics; its appearance in photoproduction was
emphasized e.g. in Ref.~\cite{Nem}. 

The reader should notice that the ZEUS data points~\cite{Priv} for $d\sigma/dt$,
$\sigma_{el}(s)$ and $B(s)$ are not plotted in Figs.~4-6. The reason 
is that these data points are not yet published by the ZEUS Collaboration.

The main result of this paper is that $J/\psi$ photoproduction is ``soft''. 
The highest-energy data~\cite{Adlo,Osaka} bring new evidence in 
favour of this observation. They show that the introduction of any ``hard''
term here is unnecessary.

This feature appears not only in the relative low, with respect to the 
``experimental data'', value of $\sigma_{el}(s)$, but even more so in its 
downwards curvature, resulting from the presence of two terms in the amplitude, 
whose interference produces the rapid rise in $\sigma_{el}(s)$ in the mid-HERA 
energy region, with a subsequent slow-down at high energies. 

More information on the possible ``hardening'' of the Pomeron with increasing
$Q^2$ may come from the extension of the present (or similar) models to 
electroproduction. However this is not easy since all (five, in our case) 
parameters will acquire some (complicated) $Q^2$ dependence, thus increasing the 
number of the fitted parameters and consequently reducing the credibility 
(confidence level) of the model and its fit. A possible solution may be 
suggested by QCD calculations of the upper vertex in Fig.~1,
$\beta_2(t)$, for virtual photons.

\begin{figure}[tb]
\begin{center}
\includegraphics[scale=0.68]{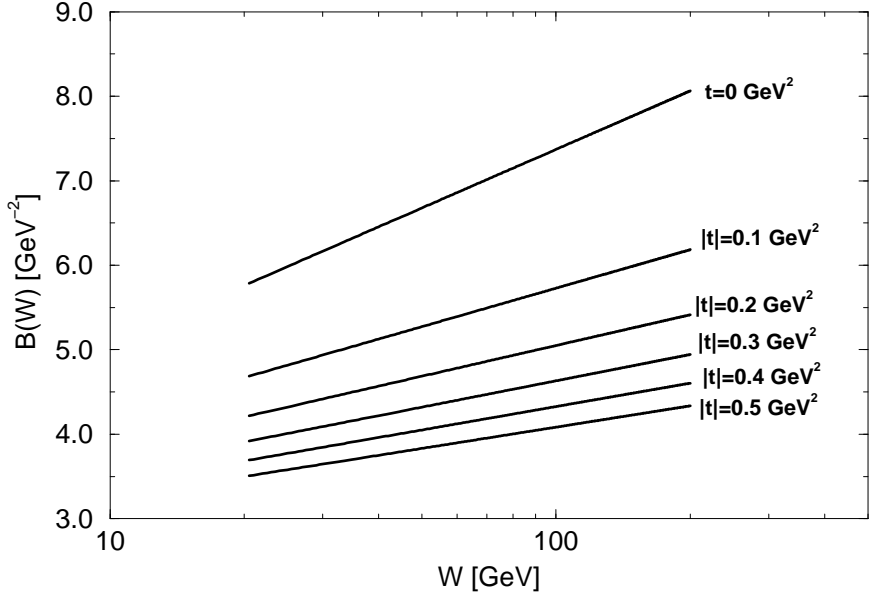}
\end{center}
\caption[]{\small Prediction from our fit to the ZEUS data for the slope $B$ 
[GeV$^{-2}$].}
\label{fig6}
\end{figure}

Finally, let us mention that the dipole Pomeron was criticized~\cite{Cudell} 
for the presence of a negative constant term in it, that -- according to 
the authors of Ref.~\cite{Cudell} -- implies a negative total cross section at low 
energies. Actually, this does not happen because of the presence of 
subasymptotic terms at low  energies, compensating the negative contribution 
($g=-5.736$, in the H1 case, $g=-4.126$, in the ZEUS case): a Pomeron daughter 
(for a purely diffractive process) and/or secondary Reggeons otherwise. 
Our fits have been made, however, well beyond the low energy region, where
subasymptotic contributions may be important.

After the completion of this work we have become aware of similar
conclusions drawn in a paper by Martynov, Predazzi and Prokudin, 
to appear~\cite{Prokudin}.

\vspace{0.5cm}
{\bf \large Acknowledgments} 

We thank M. Arneodo and A. Bruni for fruitful discussions on the ZEUS data.
L.L.J. is grateful to the Dipartimento di Fisica dell'Universit\`a della
Calabria, to the Dipartimento di Fisica dell'Universit\`a di Padova and to the 
Istituto Nazionale di Fisica Nucleare -- Sezione di Padova and Gruppo Collegato di
Cosenza, for their warm hospitality and financial support.
This work was supported in part by the Ministero Italiano dell'Universit\`a
e della Ricerca Scientifica e Tecnologica and in part by INTAS, Grant
00-00366. One of us (L.L.J.) acknowledges also the support by CRDF, Grant UP1-2119.

\vfill \eject

\end{document}